\documentclass{article}
\usepackage{spconf,amsmath,graphicx}
\usepackage{ amssymb }
\usepackage{ mathrsfs }

\usepackage{multirow}
\usepackage{graphicx}
\usepackage[table,xcdraw]{xcolor}
 
\usepackage{enumitem}
\usepackage{verbatim}
\usepackage{url}
\setlist{nosep, leftmargin=14pt}

\usepackage{mwe} 

\title{Style transfer between Microscopy and Magnetic Resonance Imaging via Generative Adversarial Network in small sample size settings}
\name{Monika Pytlarz, \quad Adrian Onicas, \quad Alessandro Crimi
}
\address{Sano – Centre for Computational Personalised Medicine}
\begin{document}

\maketitle

\begin{abstract}

Cross-modal augmentation of Magnetic Resonance Imaging (MRI) and microscopic imaging based on the same tissue samples is promising because it can allow histopathological analysis in the absence of an underlying invasive biopsy procedure. Here, we tested a method for generating microscopic histological images from MRI scans of the corpus callosum using conditional generative adversarial network (cGAN) architecture. To our knowledge, this is the first multimodal translation of the brain MRI to histological volumetric representation of the same sample. The technique was assessed by training paired image translation models taking sets of images from MRI scans and microscopy. The use of cGAN for this purpose is challenging because microscopy images are large in size and typically have low sample availability. The current work demonstrates that the framework reliably synthesizes histology images from MRI scans of corpus callosum, emphasizing the network's ability to train on high resolution histologies paired with relatively lower-resolution MRI scans. With the ultimate goal of avoiding biopsies, the proposed tool can be used for educational purposes. 
\end{abstract}
\section{Introduction}
\label{sec:intro}

The human brain is a complex system. To inspect its multi-scale organization we require several technologies, some of which are very invasive. While an anatomical representation of the brain can be easily acquired safely and non-invasively with MRI, further characterization needs histological procedures and microscopy. A biopsy is an invasive procedure to perform, therefore generating synthetic histology images complementary to MRI would be beneficial. On the other hand, comparing MRIs to the matching histology slices compensates for inevitably occurring distortions in the tissue during blocking and sectioning \cite{Mancini2020}. Histology offers great contrast at the microscopic scale due to the usage of dedicated stains that target distinct microanatomical or cytoarchitectural traits. Because of this extremely high resolution on distinct levels of magnification, a single .SVS file with one slice of a histology sample occupies almost 4 GB of data storage and requires dedicated software for the display. Considerable contrast and resolution differences between MRI and histology, often coupled with potential inhomogeneous staining and sectioning artifacts, make the alignment of these two modalities a demanding inter-modality registration problem \cite{Mancini2020}. Even though there is a growing body of literature on the topic of combining histology with other modalities, producing brain datasets with registered modalities is very laborious, so the amount of data available is still not optimal for deep learning implementations. 
The motivation for choosing GAN as a core of the framework is that GANs have produced outstanding results in image generation, image editing, and representation learning \cite{Goodfellow2020}. The concept of adversarial loss, which forces generated images to be indistinguishable from real photos, is critical to GANs' success. This loss is especially potent for image generation tasks, as this is precisely the goal that much of computer graphics seeks to optimize. GANs ability to learn image style is favorable in medical imaging – generating synthetic images was examined for medical image registration \cite{Tanner2018}, artifacts correction and increasing quality images \cite{Vey2019}, or translating MRI to computed tomography (CT) for multimodal settings \cite{Lei2019}.

The goal of the project was to apply a generative adversarial network (GAN)  able to produce synthetic histology images from MRIs with small sample size settings and to learn from images of large high resolution microscopy digital slides in the Aperio SVS format. To our knowledge, despite the recent efforts to generate different modalities with GANs in MRI \cite{han2018gan} or histology \cite{shaban2019staingan}, this is the first multimodal translation of the brain MRI stack of slices to histological volumetric representation of the same sample, which is challenging due to the completely different nature of the two modalities. 
We are not claiming that in this way, thanks to our approach MRI data can completely rule out histological data in clinical settings. Nevertheless, this work contributes to saving time and avoiding invasive histologies in some cases.  
\vspace{-1mm}
\section{Materials and methods}
\label{sec:format}
\vspace{-1mm}

\subsection{Dataset}
\label{ssec:subhead}
We use a multimodal dataset previously acquired by Tendler et al. \cite{Tendler2022}, for which ethical approval was obtained by the original authors. For this study, we selected MRIs and glial fibrillary acidic protein (GFAP) maps from the Digital Anatomist resource. Specifically, the first $b_0$ volume of the diffusion MRI (400 \textmu m isotropic at 7 T) and the corresponding registered histology (0.25 µm in-plane) from three human corpus callosum specimens were used. For more information on the data acquisition are available \cite{Tendler2022, Mollink2017}. The choice of using the $b_0$ volume was given by the fact that no traditional T1 or T2 sequence was available. It can be hypothesized that with such unavailable modalities, results should be superior.
The dataset is small and contains 5 training examples of whole slice view paired images and 3 examples of paired images for testing. Additionally, large differences in contrast, resolution, and type of details occur between MRI and histology. Making this a non-trivial case of style-transfer between images. 
\vspace{-2.5mm}
\subsection{Data preprocessing}
\label{ssec:subhead}
Due to the listed challenges, before the data is fed into the model, it has to be preprocessed (Fig. 1). Preprocessing includes downsampling of the .SVS files, registering moving MRI slices to fixed histology slices, and tiling the histology into smaller patches. We considered 2 cases:
i) downsampling whole images and performing the generation of the entire image.
ii) creating patches of smaller size and generating the resulting images also in patches. 
Codes of the pipeline are provided: https://github.com/octpsmon/Style-transfer-MRI-Histology-via-cGAN.

\textbf{SVS downsampling:} To facilitate further processing of the microscopy images and reduce memory usage while preserving the resolution, slices have been downsampled 15 times using QuPath software for digital pathology image analysis \cite{Bankhead2017}.

\textbf{Image registration:}
To align MRI and histology slices to a common coordinate system, intensity-based affine image registration was implemented using built-in MATLAB functions from the ‘Register Multimodal MRI Images’ toolbox. Sagittal slices of volumetric MRI brain representations exported from ITK-SNAP \cite{py06nimg} medical image viewer were loaded as moving images and GFAP maps as fixed. Downsampled GFAP maps were of size around 6000x10000 and MRIs - 98x128. The mean Dice score evaluating the obtained alignment between fixed and moving is 0.93.

\vspace{-1mm}

\begin{figure}[hb]
    \begin{minipage}[Fig. 1]{1.0\linewidth}
      \centering
      \centerline{\includegraphics[width=9.6cm]{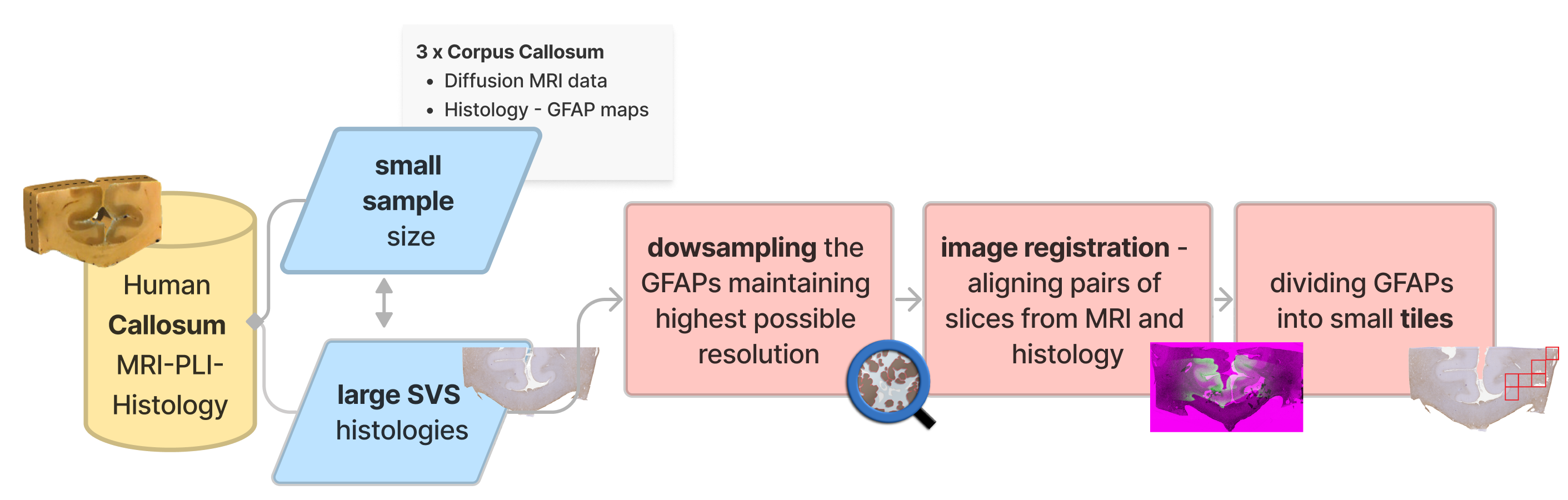}}
\end{minipage}
\caption{Image preprocessing pipeline. Whole slide histology images were downsampled, registered with MRIs, and tiled.}
\label{fig:res}
\end{figure}

\textbf{Splitting into patches:} Whole slide images are too large to fit on a GPU at once; instead, they are usually divided into smaller patches for training the deep learning model. Images of the order of magnitude 10000x10000 have been cut into 1024x1024 and 256x256 non-overlapping tiles using ImageJ plugin SlideJ \cite{DellaMea2017}. Then the data was fed into the model within 3 experiments: 1) images as downsampled and scaled whole slice views 4096x4096 (5 paired images for training, 3 for testing) 2) slices as groups of 1024x1024 patches (385 for training, 248 for testing), 3) slices as groups of 256x256 patches (5722 for training, 3819 for testing).
\vspace{-2mm}
\subsection{Network architecture}
\label{ssec:subhead}
Image-to-image translation aims at learning the mapping between the input image and an output image using a set of registered pairs of images. Given the nature of our dataset, a cGAN model was trained on paired data (each histology section paired and registered with MRI slices).
We adapted the publicly available PyTorch implementation of pix2pix, a cGAN model, by Isola et al. \cite{Isola2017} to translate MRIs to histology, incorporating an image registration preprocessing step and hyperparameter tuning.
The model has two architectures, one for the generator and the other for the discriminator, specifically encoder-decoder U-net and patchGAN. Discriminator's patchGAN architecture includes a number of transpose convolutional blocks. It examines an NxN section of an image to determine whether it is real or fake \cite{Isola2017}.
Conditional GANs learn the mapping from an observed image x and a random noise vector z to y, G: $x, z \rightarrow y$. G - generator - is trained to produce outputs that cannot be distinguished from real images by the discriminator. A discriminator, D, is adversarially trained to do as well as possible at spotting the generator's "fakes". Load sizes chosen for 3 training experiments were: 4096, 1024, and 256. The adversarial loss for the conditional GAN is defined as:
\vspace{-2mm}
\begin{equation}
\begin{array}{l}
\mathcal{L}_{\text{cGAN}}(G, D) = \\
\mathbb{E}_{x, y}[\log D(x, y)] + \mathbb{E}_{x, z}[\log(1 - D(x, G(x, z)))],
\end{array}
\end{equation}
\vspace{-1mm}
  where $ D $ denotes the discriminator corresponding to the generator $ G $. Then, using jointly the L1 distance and considering the $z$ random noise vector,  
 \[
\mathcal{L}_{L1}(G) = \mathbb{E}_{x, y, z} \left[ \| y - G(x, z) \|_1 \right]
\]
leads to the final objective:
\vspace{-2mm}
\begin{equation}
G^* = \arg\min_G \max_D \mathcal{L}_{\text{cGAN}}(G, D) + \lambda \mathcal{L}_{L1}(G).
\end{equation}
\vspace{-1mm}

\vspace{-2mm}
\subsection{Hyperparameter tuning}
\label{ssec:subhead}
Lucic and colleagues \cite{lucic2018gans} observed that GAN training is incredibly sensitive to hyperparameter settings. Therefore, several hyperparameters have been tested using a randomized search approach to select the ones which give the best results of training on the whole slide view, based on the calculated loss function scores as endpoints. The applied L1 loss function measures the mean absolute difference between the generated image and the target image, helping to enforce pixel-level similarity between the generated image and the target image. Default settings are U-net 256 blocks as architecture network, cross-entropy loss function, learning rate of value 2x10-4, beta1 momentum term of adam optimizer: 0.5, number of epochs: 100, weight for L1 loss 'lambda$\_$L1': 100. The following parameters were varying within given options: type of GAN objective (cross-entropy - 'vanilla', least squares - 'lsgan', Wasserstein distance - 'wgangp') the number of epochs with the initial learning rate, learning rate (2x10-4, 2x10-5), beta1 - momentum term of adam optimizer (0.4, 0.5, 0.8), lambda$\_$L1 (10, 50, 100). 

Cross-entropy GAN loss is a binary classification loss \cite{NIPS2014_5ca3e9b1} used to train the discriminator in GAN. The difference between the true label and the predicted label of a classification model is measured. In the context of a GAN, the true label is 1 for real data and 0 for generated data, and the predicted label is the output of the discriminator. The binary cross-entropy loss is calculated for each data point (real or generated) separately and then averaged over the entire batch of data. 

Least squares loss function has been adopted from the Least Squares Generative Adversarial Networks (LSGANs) \cite{9732473} where the least squares loss function replaces the binary cross-entropy. In LSGANs, the generator is trained to minimize a least squares loss function, which measures the difference between the discriminator's output on the generated data and a continuous target value. The LSGAN loss function is designed to overcome some of the instability issues associated with traditional GANs, such as mode collapse and slow convergence.

The Wasserstein distance \cite{pmlr-v70-arjovsky17a}, also known as the Earth Mover's Distance, measures the minimum energy cost of transforming one distribution into another. Wasserstein GANs, which use the Wasserstein distance as a loss function, have been shown to be effective in generating high-quality images that are similar to the real images in terms of their distribution. Wasserstein Gradient Penalty Loss \cite{gulrajani2017improved}, or WGAN-GP Loss, is a loss used for generative adversarial networks that augment the Wasserstein loss with a gradient norm penalty for random samples to achieve Lipschitz continuity. A Lipschitz continuous function is a mathematical concept that describes a function whose rate of change is bounded. Encouraging the discriminator in a GAN to have Lipschitz continuous gradients is beneficial because it helps to stabilize the training process and prevent mode collapse.

\vspace{-2mm}
\subsection{Evaluation}
\label{ssec:subhead}
Apart from the L1 loss function generated images have been evaluated using two additional metrics: Frechet Inception Distance (FID) \cite{Heusel2017} and Learned Perceptual Image Patch Similarity (LPIPS) \cite{Zhang2018-tt}. They provide a more comprehensive assessment of the quality of the generated images since they measure perceptual similarity rather than the statistical similarity between generated and real images.
FID score \cite{Heusel2017}, assesses the quality of generated images comparing the similarity of two datasets. It is applied to compute the Frechet inception distance using the inception network to measure the distance between the generated image distribution and the real image distribution. The FID metric calculates the maximum entropy distribution for a given mean and covariance \cite{lopes2021creating}.
LPIPS \cite{Zhang2018-tt} measures the distance between the feature representations of real and generated images. Using a pre-trained network it extracts image features and calculates the Euclidean distance between the feature vectors. LPIPS's authors believe that perceptual similarity is not a distinct function of its own, but rather the result of visual representations tuned to predict important world structure. Representations that perform well in semantic prediction tasks are also those in which Euclidean distance predicts perceptual similarity judgments very well.

\vspace{-2mm}
\begin{figure}[!htb]

\begin{minipage}[Fig. 2]{\linewidth}
  \centering
  \centerline{\includegraphics[width=\linewidth]{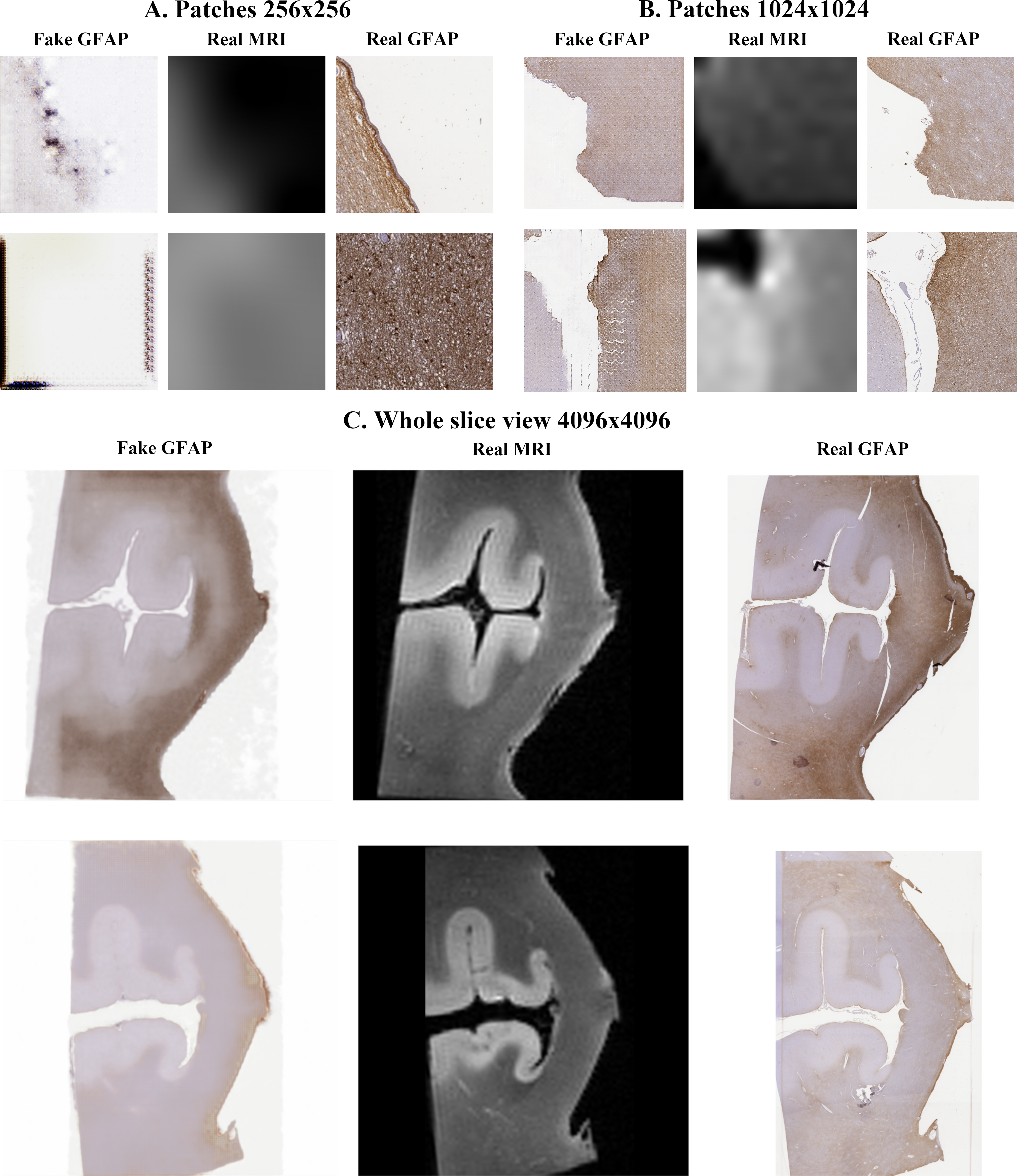}} 

\end{minipage}
\caption{Qualitative evaluation. Sample images from 3 experiments, 2 for each of them, one example is included in one row. The top left and right rows are for the patches 
experiments, and the bottom row is for the full slice experiment.}
\label{fig:res}
\end{figure}
\vspace{-8mm}
\section{Results}
\vspace{-2mm}
\label{sec:pagestyle}
Model was trained on the supercomputer of over 3.5 PFlops for the CPU parts, and over 500 TFlops for the GPU parts. Training on whole slices took approximately 3h, on 1024x1024 patches - 4h, and on 256x256 patches - 8h. In Table 1. there are presented calculated FID and LPIPS scores comparing generated histologies to real (ground truth) histologies and also generated histologies to real MRIs. Sample results for qualitative assessment are provided in Figure 2.
Two best models selected from hyperparameter tuning are of subsequent combinations of parameters: 1) U-net 256 blocks network, least squares GAN objective, 100 epochs, beta1: 0.4, learning rate - 2x10-4, lambda$\_$L1 - 10, 2) U-net 256 blocks network, least squares GAN objective, 100 epochs, beta1 - 0.8, learning rate - 2x10-4, lambda$\_$L1 - 50.

\begin{table}[]
\centering
\resizebox{\columnwidth}{!}{%
\begin{tabular}{llllll}
\hline
 &
  \multicolumn{2}{c}{\textbf{Patches}} &
  \multicolumn{3}{c}{\textbf{Whole slice view (4096x4096)}} \\ \cline{2-6}
\multirow{-2}{*}{\textbf{Comparison}} &
  \multicolumn{1}{c}{\textbf{256x256}} &
  \multicolumn{1}{c}{\textbf{1024x1024}} &
  \multicolumn{1}{c}{\textbf{Default model}} &
  \multicolumn{1}{c}{\textbf{Best model}} &
  \multicolumn{1}{c}{\textbf{2nd best model}} \\ \hline
\multicolumn{6}{c}{\cellcolor[HTML]{EFEFEF}\textit{\textbf{FID score – train set}}} \\ \hline
\begin{tabular}[c]{@{}l@{}}Generated histology\\ vs real histology\end{tabular} &
  \textbf{138.990} &
  \textbf{154.650} &
  511.090 &
  414.554 &
  496.001 \\ \hline
\begin{tabular}[c]{@{}l@{}}Generated histology\\ vs real MRI\end{tabular} &
  \textbf{449.312} &
  \textbf{347.470} &
  616.730 &
  557.848 &
  607.187 \\ \hline
\multicolumn{6}{c}{\cellcolor[HTML]{EFEFEF}\textit{\textbf{FID score – test set}}} \\ \hline
\begin{tabular}[c]{@{}l@{}}Generated histology\\ vs real histology\end{tabular} &
  \textbf{163.710} &
  \textbf{235.570} &
  630.390 &
  511.664 &
  549.092 \\ \hline
\begin{tabular}[c]{@{}l@{}}Generated histology\\ vs real MRI\end{tabular} &
  449.311 &
  400.050 &
  \textbf{387.340} &
  \textbf{398.535} &
  \text{425.434} \\ \hline
\multicolumn{6}{c}{\cellcolor[HTML]{EFEFEF}\textit{\textbf{LPIPS score – train set}}} \\ \hline
\begin{tabular}[c]{@{}l@{}}Generated histology\\ vs real histology\end{tabular} &
  \text{0.488} &
  \textbf{0.369} &
  0.546 &
  \cellcolor[HTML]{FFFFFF}\textbf{0.416} &
  0.435 \\ \hline
\begin{tabular}[c]{@{}l@{}}Generated histology\\ vs real mri\end{tabular} &
  0.977 &
  \text{1.081} &
  \textbf{0.533} &
  0.698 &
  \textbf{0.619} \\ \hline
\multicolumn{6}{c}{\cellcolor[HTML]{EFEFEF}\textit{\textbf{LPIPS score – test set}}} \\ \hline
\begin{tabular}[c]{@{}l@{}}Generated histology\\ vs real histology\end{tabular} &
  \textbf{0.341} &
  0.624 &
  0.589 &
  \textbf{0.485} &
  \text{0.818} \\ \hline
\begin{tabular}[c]{@{}l@{}}Generated histology\\ vs real MRI\end{tabular} &
  0.856 &
  1.280 &
  \textbf{0.811} &
  0.832 &
  \textbf{0.500} \\ \hline
\end{tabular}
}
\caption{Mean FID and LPIPS scores – comparison of generated to real histologies (ground truth) and to real MRIs. Hyperparameters were tuned on the whole slice view. Bold indicates 2 lowest FIDs and LPIPSs of training and test sets for each comparison.}
\label{tab:my-table}
\end{table}

\vspace{-2mm}
\section{Discussion}
\label{sec:typestyle}
\vspace{-2mm}
Calibrating the training with random search was beneficial. The best acquired model as a loss function uses least squares, which corresponds with evidence that approaches adopted from LSGAN produce high-quality results in various image generation tasks \cite{AbuSrhan2022, jitkrittum2018informative, Wang2021}.
The FID scores obtained by models trained on patches are better than the ones for whole slice view. However, during the inference in the particular setting that the project was implemented for - which is generating histology directly from the MRI - the best quantitative result was reached for the whole slice view (in relation to the default and tuned model). Jointly, LPIPS score values show that the optimized model trained on whole slice view works the best in translating MRI to histology given unseen data. LPIPS metric reflects more the positive impact of the hyperparameters optimization to model performance.

In contrast to metrics directly comparing images pixel by pixel, perceptual FID, and LPIPS trained on deep features tend to mimic human perception of similarity in images, but still, when evaluating generative adversarial networks, the qualitative assessment of the results is needed. Sample output images show that when histology is produced from the whole slice view, the borders of the tissue are preserved and the structures are distinguishable. Almost similarly the features of the MRIs are translated from 1024x1024 patches. Additionally, the texture of the synthetic histology is more detailed. The model trained on the smallest patches performs the smallest quality images among 3 experiments. This size of tiles seems to be too small to provide enough context - especially from the MRI image, which intrinsically has a significantly lower resolution. Among the stack of the 256x256 tiles, the generated outputs are very light, poorly detailed, and rarely reflect even the sharp lines visible in the MRI.

The smallest crops used to train the network lack relevant information, and thus the training may fail: both the generator and the discriminator require information to process and may encounter issues if that information is not available. Even if the training is successful, when stitching all the different crops of a very high resolution image, the stylistic contribution of each small translated image can be insufficient for the entire high resolution image. The generator has no knowledge of the context of the entire high definition (HD) image and is only exposed to the lower resolution 256x256 crops. Giving the generator some encoded context about the entire images can certainly broaden the technique's range of applications, offering complex context-aware HD image translations. Since the generated 1024x1024 patches are satisfactory, and this size parameter allows to preservation of relevant elements of medical image, it is possible that joining back the tiles with respect to initially encoded coordinates will be already a sufficient solution. 
Future works include the comparison of GANs with diffusion models \cite{dhariwal2021diffusion}, and estimation of tractography from both types of data using structural tensors \cite{khan20153d}.

\vspace{-5mm}
\vspace{-2mm}
\section{Conclusion}
\label{sec:majhead}
\vspace{-2mm}
We proposed a deep learning-based approach to synthesize a histology image directly from a brain MRI. The demonstrated method incorporates a framework of cGAN. We proved the model is capable of reliably learning the style from one modality and translating it to another, even when they are particularly different as histology and MRI. Preliminary results were promising, showing the network's ability to train on high resolution histologies paired with relatively low-resolution MRI modality.
It is probably too early to be included in clinical workflow avoiding histology, as this will require further improvements. Nevertheless, with currently accomplished scores, the method can be reliably used for educational purposes saving time to pathology laboratories.
\vspace{-4mm}
\section{Acknowledgments}
\label{sec:acknowledgments}
 \vspace{-1mm}
This project has received funding from the European Union’s Horizon 2020 research and innovation programme under grant agreement No 857533 and from the International Research Agendas Programme of the Foundation for Polish Science No MAB PLUS/2019/13.

{\small
\bibliographystyle{IEEEbib-abbrev}
\bibliography{refs.bib}

\begin{thebibliography}{10}

\bibitem{Mancini2020}
M. Mancini, A. Casamitjana, L. Peter, E. Robinson, S. Crampsie, D.~L. Thomas, J.~L. Holton, Z. Jaunmuktane, and J.~E. Iglesias,
\newblock ``{A multimodal computational pipeline for 3D histology of the human brain},''
\newblock {\em Scientific Reports}, vol. 10, no. 1, pp. 1--21, 2020.

\bibitem{Goodfellow2020}
I. Goodfellow, J. Pouget-Abadie, M. Mirza, B. Xu, D. Warde-Farley, S. Ozair, A. Courville, and Y. Bengio,
\newblock ``{Generative adversarial networks},''
\newblock {\em Communications of the ACM}, vol. 63, no. 11, pp. 139--144, 2020.

\bibitem{Tanner2018}
C. Tanner, F. Ozdemir, R. Profanter, V. Vishnevsky, E. Konukoglu, and O. Goksel,
\newblock ``{Generative Adversarial Networks for MR-CT Deformable Image Registration},''
\newblock pp. 1--11, 2018.

\bibitem{Vey2019}
B.~L. Vey, J.~W. Gichoya, A. Prater, and C.~M. Hawkins,
\newblock ``{The Role of Generative Adversarial Networks in Radiation Reduction and Artifact Correction in Medical Imaging},''
\newblock {\em Journal of the American College of Radiology}, vol. 16, no. 9, pp. 1273--1278, 2019.

\bibitem{Lei2019}
Y. Lei, J. Harms, T. Wang, Y. Liu, H.~K. Shu, A.~B. Jani, W.~J. Curran, H. Mao, T. Liu, and X. Yang,
\newblock ``{MRI-only based synthetic CT generation using dense cycle consistent generative adversarial networks},''
\newblock {\em Medical Physics}, vol. 46, no. 8, pp. 3565--3581, 2019.

\bibitem{han2018gan}
C. Han, H. Hayashi, L. Rundo, R. Araki, W. Shimoda, S. Muramatsu, Y. Furukawa, G. Mauri, and H. Nakayama,
\newblock ``Gan-based synthetic brain mr image generation,''
\newblock in {\em 2018 IEEE 15th international symposium on biomedical imaging (ISBI 2018)}. IEEE, 2018, pp. 734--738.

\bibitem{shaban2019staingan}
M.~T. Shaban, C. Baur, N. Navab, and S. Albarqouni,
\newblock ``Staingan: Stain style transfer for digital histological images,''
\newblock in {\em 2019 Ieee 16th international symposium on biomedical imaging (Isbi 2019)}. IEEE, 2019, pp. 953--956.

\bibitem{Tendler2022}
B.~C. Tendler, T. Hanayik, and O.~e.~a. Ansorge,
\newblock ``{The Digital Brain Bank, an open access platform for post-mortem datasets},''
\newblock {\em eLife}, vol. 11, pp. 1--35, 2022.

\bibitem{Mollink2017}
J. Mollink, M. Kleinnijenhuis, A.~M. van {Cappellen van Walsum}, and e.~a. Sotiropoulos,
\newblock ``{Evaluating fibre orientation dispersion in white matter: Comparison of diffusion MRI, histology and polarized light imaging},''
\newblock {\em NeuroImage}, vol. 157, pp. 561--574, 2017.

\bibitem{Bankhead2017}
P. Bankhead, M.~B. Loughrey, J.~A. Fern{\'{a}}ndez, and D. et~al.,
\newblock ``{QuPath: Open source software for digital pathology image analysis},''
\newblock {\em Scientific Reports}, vol. 7, no. 1, pp. 1--7, 2017.

\bibitem{py06nimg}
P.~A. Yushkevich, J. Piven, H. Cody~Hazlett, R. Gimpel~Smith, S. Ho, J.~C. Gee, and G. Gerig,
\newblock ``User-guided {3D} active contour segmentation of anatomical structures: Significantly improved efficiency and reliability,''
\newblock {\em Neuroimage}, vol. 31, no. 3, pp. 1116--1128, 2006.

\bibitem{DellaMea2017}
V. {Della Mea}, G.~L. Baroni, D. Pilutti, and C. {Di Loreto},
\newblock ``{SlideJ: An ImageJ plugin for automated processing of whole slide images},''
\newblock {\em PLoS ONE}, vol. 12, no. 7, pp. 1--9, 2017.

\bibitem{Isola2017}
P. Isola, J.~Y. Zhu, T. Zhou, and A.~A. Efros,
\newblock ``{Image-to-image translation with conditional adversarial networks},''
\newblock {\em Proceedings - 30th IEEE Conference on Computer Vision and Pattern Recognition, CVPR 2017}, vol. 2017-Janua, pp. 5967--5976, 2017.

\bibitem{lucic2018gans}
M. Lucic, K. Kurach, M. Michalski, S. Gelly, and O. Bousquet,
\newblock ``Are gans created equal? a large-scale study,''
\newblock {\em Advances in neural information processing systems}, vol. 31, 2018.

\bibitem{NIPS2014_5ca3e9b1}
I. Goodfellow, J. Pouget-Abadie, M. Mirza, B. Xu, D. Warde-Farley, S. Ozair, A. Courville, and Y. Bengio,
\newblock ``Generative adversarial nets,''
\newblock in {\em Advances in Neural Information Processing Systems}, Z. Ghahramani, M. Welling, C. Cortes, N. Lawrence, and K. Weinberger, Eds. 2014, vol.~27, Curran Associates, Inc.

\bibitem{9732473}
C.-K. Lee, Y.-J. Cheon, and W.-Y. Hwang,
\newblock ``Least squares generative adversarial networks-based anomaly detection,''
\newblock {\em IEEE Access}, vol. 10, pp. 26920--26930, 2022.

\bibitem{pmlr-v70-arjovsky17a}
M. Arjovsky, S. Chintala, and L. Bottou,
\newblock ``{W}asserstein generative adversarial networks,''
\newblock in {\em Proceedings of the 34th International Conference on Machine Learning}, D. Precup and Y.~W. Teh, Eds. 06--11 Aug 2017, vol.~70 of {\em Proceedings of Machine Learning Research}, pp. 214--223, PMLR.

\bibitem{gulrajani2017improved}
I. Gulrajani, F. Ahmed, M. Arjovsky, V. Dumoulin, and A.~C. Courville,
\newblock ``Improved training of wasserstein gans,''
\newblock {\em Advances in neural information processing systems}, vol. 30, 2017.

\bibitem{Heusel2017}
M. Heusel, H. Ramsauer, T. Unterthiner, B. Nessler, and S. Hochreiter,
\newblock ``{GANs trained by a two time-scale update rule converge to a local Nash equilibrium},''
\newblock {\em Advances in Neural Information Processing Systems}, vol. 2017-Decem, no. Nips, pp. 6627--6638, 2017.

\bibitem{Zhang2018-tt}
R. Zhang, P. Isola, A.~A. Efros, E. Shechtman, and O. Wang,
\newblock ``The unreasonable effectiveness of deep features as a perceptual metric,''
\newblock in {\em 2018 {IEEE/CVF} Conference on Computer Vision and Pattern Recognition}. June 2018, IEEE.

\bibitem{lopes2021creating}
D.~J.~V. Lopes, G.~F. Monti, G.~W. Burgreen, J.~C. Moulin, G. dos Santos~Bobadilha, E.~D. Entsminger, and R.~F. Oliveira,
\newblock ``Creating high-resolution microscopic cross-section images of hardwood species using generative adversarial networks,''
\newblock {\em Frontiers in Plant Science}, vol. 12, pp. 760139, 2021.

\bibitem{AbuSrhan2022}
A. Abu-Srhan, M.~A. Abushariah, and O.~S. Al-Kadi,
\newblock ``The effect of loss function on conditional generative adversarial networks,''
\newblock {\em Journal of King Saud University - Computer and Information Sciences}, vol. 34, no. 9, pp. 6977--6988, Oct. 2022.

\bibitem{jitkrittum2018informative}
W. Jitkrittum, H. Kanagawa, P. Sangkloy, J. Hays, B. Sch{\"o}lkopf, and A. Gretton,
\newblock ``Informative features for model comparison,''
\newblock {\em Advances in Neural Information Processing Systems}, vol. 31, 2018.

\bibitem{Wang2021}
J. Wang, X. Chang, Y. Wang, R.~J. Rodr{\'{\i}}guez, and J. Zhang,
\newblock ``{LSGAN}-{AT}: enhancing malware detector robustness against adversarial examples,''
\newblock {\em Cybersecurity}, vol. 4, no. 1, Dec. 2021.

\bibitem{dhariwal2021diffusion}
P. Dhariwal and A. Nichol,
\newblock ``Diffusion models beat gans on image synthesis,''
\newblock {\em Advances in Neural Information Processing Systems}, vol. 34, pp. 8780--8794, 2021.

\bibitem{khan20153d}
A.~R. Khan, A. Cornea, L.~A. Leigland, S.~G. Kohama, S.~N. Jespersen, and C.~D. Kroenke,
\newblock ``3d structure tensor analysis of light microscopy data for validating diffusion mri,''
\newblock {\em Neuroimage}, vol. 111, pp. 192--203, 2015.

\end{thebibliography}
}
\end{document}